\documentclass{llncs}

\usepackage{makeidx}         
\usepackage{graphicx}        
\usepackage{multicol}        
\usepackage[bottom]{footmisc}

\makeindex             

\usepackage{amsmath}
\usepackage{amsfonts}
\usepackage{graphics}
 \usepackage{amssymb}
 \usepackage{bbm}
 \usepackage{times}
\usepackage{amstext}
\usepackage{theorem}

\begin{document}

\newcommand{\sometime}{\lozenge}
\newcommand{\last}{\bigodot}
\newcommand{\alwaysP}{\blacksquare}
\newcommand{\sometimesP}{\bla\documentclass{llncs}cklozenge}
\newcommand{\LTLL}{$LTL_{\lambda=}\;$}

\newcommand{\always}{\raisebox{-.2ex}{
			   \mbox{\unitlength=0.9ex
			   \begin{picture}(2,2)
			   \linethickness{0.06ex}
			   \put(0,0){\line(1,0){2}}
			   \put(0,2){\line(1,0){2}}
			   \put(0,0){\line(0,1){2}}
			   \put(2,0){\line(0,1){2}}
			   \end{picture}}}
		      \,}

\newcommand{\next}{\!\raisebox{-.2ex}{ 
			\mbox{\unitlength=0.9ex
			\begin{picture}(2,2)
			\linethickness{0.06ex}
			\put(1,1){\circle{2}} 
			\end{picture}}}       
			\,}

\newcommand{\until}{\mathcal{U}}

\newcommand{\while}{\mathcal{W}}

\newcommand{\since}{\mathcal{S}}

\newcommand{\union}{\cup}
\newcommand{\auf}{\left\langle}
\newcommand{\zu}{\right\rangle}
\newcommand{\bydef}{\rightleftharpoons}   

\newcommand{\Qu}{\mathbb Q}
\newcommand{\Zed}{\mathbb Z}
\newcommand{\Complex}{\mathbb C} 
\newcommand{\Real}{\mathbb R}

\newcommand{\uP}{\mathcal{U}}
\newcommand{\tP}{\mathcal{T}}
\newcommand{\eP}{\mathcal{E}}
\newcommand{\sP}{\mathcal{S}}
\newcommand{\iP}{\mathcal{I}}
\newcommand{\pP}{\mathcal{P}}
\newcommand{\mU}{\mathcal{U}}
\newcommand{\mT}{\mathcal{T}}
\newcommand{\mS}{\mathcal{S}}
\newcommand{\mP}{\mathcal{P}}
\newcommand{\mL}{\mathcal{L}}
\newcommand{\mA}{\mathcal{A}}
\newcommand{\mB}{\mathcal{B}}
\newcommand{\mC}{\mathcal{C}}
\newcommand{\mF}{\mathcal{F}}
\newcommand{\mG}{\mathcal{G}}
\newcommand{\mH}{\mathcal{H}}
\newcommand{\mM}{\mathcal{M}}
\newcommand{\mD}{\mathcal{D}}
\newcommand{\gD}{\mathfrak{D}}
\newcommand{\gI}{\mathfrak{I}}
\newcommand{\mI}{\mathcal{I}}
\newcommand{\gM}{\mathfrak{M}}
\newcommand{\ga}{\mathfrak{a}}
\newcommand{\gma}{\models^{\mathfrak{a}}}
\newcommand{\gmaa}{\models^{\mathfrak{a}'}}
\newcommand{\gA}{\mathfrak{V}}
\newcommand{\gB}{\mathfrak{B}}
\newcommand{\gF}{\mathfrak{F}}
\newcommand{\gC}{\mathfrak{C}}
\newcommand{\gn}{\mathfrak{n}}

\title{A Logic with Temporally Accessible Iteration}
\author{Alexei Lisitsa\inst{1}}
\institute{Department of Computer Science, the University of Liverpool,\\ 
Ashton Building,Ashton Street,\\ 
Liverpool, L69 7ZF, U.K.\\
\texttt{alexei@csc.liv.ac.uk}
}
%
%
\maketitle

\begin{abstract}
Deficiency in expressive power of the first-order logic has led to developing  its numerous
extensions by fixed point operators, such as Least Fixed-Point (LFP), 
inflationary fixed-point (IFP), partial fixed-point (PFP), etc. These logics have
been extensively studied in finite model theory, database theory, descriptive
complexity. In this paper we introduce unifying framework, the logic with
iteration operator, in which iteration steps may be accessed by temporal logic formulae. 
We show that proposed logic  FO+TAI subsumes all mentioned fixed
point extensions as well as many other fixed point logics as natural fragments. On the other hand we 
show that over finite structures  FO+TAI is no more expressive than FO+PFP. Further we show that adding the same machinery to the logic of monotone inductions (FO+LFP) does not increase its expressive power either. 
\end{abstract}


\section{Introduction}

Probably one of the earliest proposals to extend logic with inductive constructs can be 
found in the Wittgenstein's famous  \emph{Tractatus
Logico-Philosophicus} \cite{Wittgenstein::1921}
\begin{quote}
\emph{4.1273  If we want to express in conceptual notation the general proposition `b is a successor of a', then 
we require an expression for the general term of the series of forms 
\begin{center}
$aRb$\\
$(\exists x); aRx.XRb$\\
$(\exists x,y):aRx.xRy.yRb$\\
$\ldots$
\end{center}
}
  \end{quote}

Implicitly Wittgenstein admitted insufficient expressive power of the (first-order) predicate logic and proposed an extension  which in modern terms we can call first-order logic augmented with the transitive closure operator (FO+TC).
Transitive closure  is a particular case of more general inductive operators, which have  extensively studied in recursion theory and its generalizations 
\cite{Post43,Mos74,Acz77}.

Special role logics with inductive operators play in foundations of computer science. Logic languages with fixed point constructs serve  theoretical models of query languages in \emph{database theory} and, when considered over linearly ordered finite structures,  are used to characterize computational complexity classes within \emph{descriptive complexity} theory\cite{Vardi82,Immerman86,Descriptive99}. The relationships between fixed point logics and complexity have many interesting aspects - the logics reflect faithfully computations over structures and this led to formulation of a new notion of \emph{relational complexity} \cite{AVV97}. On the other hand, tantalizing open problems in computational complexity can be formulated in logical terms, for example PTIME = PSPACE if and only if logics with least fixed point and partial fixed points have the same expressive power  over classes of finite models 
\cite{AbiteboulVianu}.  
In other direction, modal logic with fixed points, $\mu$-calculus,  is one of the unifying formalisms used in the research on model checking and  verification \cite{EA96}. 
Not necessarily monotone inductive definitions also appear in the research on semantics of logic programming\cite{Fitting}, in formalization of reasoning \cite{DT07}   and in the revision theory\cite{Lowe:revision}.

In this paper we propose a simple mechanism allowing to "internalize" various variants of the inductive definitions within a single logic. Semantics of fixed-point operators is usually defined by using an iteration, more precisely 
in terms of "to what iteration converge".  We suggest to look on the iteration process itself and augment the logic with an 
access to the iteration stages via temporal formulae. As a result we get a logic FO+TAI (temporally accessible iteration)  which naturally subsumes many (virtually all deterministic variants of ) inductive logics, including logics with least fixed point, inflationary fixed 
point, variants of partial fixed points, as well as logics with anti-monotone and non-monotone inductions. 

We present the semantics of FO+TAI for finite structures only. The case of infinite structures requires considering \emph{transfinite iterations} and temporal access to the iteration stages would need a variant of temporal logic over ordinals (e.g. \cite{Ord}). This case requires further investigations and will be treated elsewhere. 

We show by translations that over finite structures FO+TAI is not less expressive than all mentioned inductive logics and at the same time it is no more expressive than FO+PFP. Further, we show that adding the same machinery to the logic of monotone inductions (FO+LFP) does not increase its expressive power either.

The paper is organized as follows. In the next section we introduce classical fixed-point logics and first-order temporal logics. Based on that in the Section~\ref{FO+TAI} we define the logic FO+TAI. In Section~\ref{sec:others} we demonstrate how to define in FO+TAI classical inductive constructs. In Section~\ref{sec:ID}  it is shown tha FO+TAI subsusmes the logic of non-monotone induction FO+ID. In Section~\ref{sec:expr} we consider expressive power FO+TAI and its monotone fragment. Section~\ref{sec:conclusion} concludes the paper.

\section{Preliminaries}

\subsection{Fixed point extensions of first-order logic}
We start with the short review of inductive definability, which will set up a context in which logics with temporally 
accessible iteration naturally appear. In this paper we will mainly deal with definability over (classes of) 
finite structures, so unless otherwise stated all structures are assumed to be finite.   

Let $\varphi(R,\bar{x})$ is a first-order formula, where $R$ is a relation symbol of some arity $n$ and $\bar{x}$ 
is a tuple of individual variables of 
the length  $n$ (the same as the arity of $R$). Consider a structure ${\cal M}$ with the 
domain $M$, interpreting all symbols in 
$\varphi$ except $R$ and $\bar{x}$.  Then one  can consider a map 
$  \Phi_{\varphi}: 2^{M^{k}} \rightarrow 2^{M^{k}}$, i.e mapping $k$-ary relations over $M$ to $k$-ary
relations over $M$  defined by $\varphi(R, \bar{x})$ as follows: 

\[
\Phi_{\varphi}(P) = \{\bar{a} \mid ({\cal M}, \models \varphi(P, \bar{a}) \}
\]

Various  fixed-point constructions may then be defined. If operator
$\Phi_{\varphi}$ is monotone then by classical Knaster-Tarski theorem \cite{Tarski55} it has a least fixed-point,
that is the least relation $R$, such that $R(\bar{x}) \leftrightarrow
\varphi(R,\bar{x})$ holds. This least fixed-point $R^{\infty}$ can be obtaned as a limit of 
the following iteration:  

\begin{itemize}
\item $R_{0} = \emptyset$
\item $R_{i+1} = \Phi(R_{i})$
\end{itemize}

Over finite structures this iteration stabilizes on some finite step $n \ge 0$:  
$R_{n+1} = R_{n}$. 
Simple syntactical property of $\varphi(R,\bar{x})$ which guarantees monotonicity of $\Phi_{\varphi}$ is  that this formula is \emph{positive} in $R$.

\emph{Inflationary fixed point} of a not necessary monotone operator  $\Phi$  is defined as the 
limit of the following iteration:

\begin{itemize}
\item $R_{0} = \emptyset$
\item $R_{i+1} = \Phi(R_{i}) \cup R_{i}$
\end{itemize}

The inflationary fixed point   exists for an arbitrary operator and over finite
structures the above iteration reaches it at some finite step.

\emph{Partial fixed point} of an operator $\Phi$ defined by an arbitrary  
formula  $\varphi(R, \bar{x})$ is defined as follows. Consider the iteration:

\begin{itemize}
\item  $R_{0} = \emptyset$
\item  $R_{i+1} = \Phi(R_{i})$ 
\end{itemize}  
 
Partial fixed point of $\Phi$ is a fixed point (limit)  of the iteration 
(if it exists) and empty set otherwise.

Aiming to resolve difficulties in the definition of semantics of partial fixed point operator over infinite structures in \cite{Kreutzer}  an alternative \emph{general} semantics for such an operator has been proposed. We will discuss it later in 
\ref{subsec:gen}.

Let IND is one of the above fixed point operators
(LFP, IFP, PFP or PFPgen) then the syntax of logic FO+IND extends the standard syntax of 
first-order logic with  the following construct. Let $\varphi(R,\bar{x})$ be a formula with free individual variables
$\bar{x} = x_{1}, \ldots, x_{k}$ and free predicate variable $R$. For the case IND $\equiv$ LFP we additionally require that $\varphi(R,\bar{x})$ is positive in $R$. Then $\rho := [IND_{R, \bar{x}} \varphi]\bar{t}$ is
also formula. Free variables of $\rho$ are free variables occurring in $\varphi$ and $t$ other than $\bar{x}$. 
Semantics of such formula $\rho$ is read then as follows: an interpretation of tuple of terms $\bar{t}$ belong to the relation
which is a fixed point of the operator $\Phi_{\varphi}$ of the corresponding type IND (i.e. least, inflationary, partial, or generalized partial fixed point, for IND $\equiv$ LFP, IFP, PFP, genPFP, respectively.) 

Usually the above logics defined in a way allowing also simultaneous inductive definitions, i.e  the formulae of the
 form $[ IND \;\; R_{i} : S] \bar{t}$ where

$$
S:= 
\left\{
    \begin{aligned}
     R_{1}(\bar{x_{1}}) \leftarrow \varphi_{1}(R_{1}, \ldots, R_{k}, \bar{x_{1}})\\
     \vdots \\
     R_{k}(\bar{x_{k}})  \leftarrow \varphi_{1}(R_{1}, \ldots, R_{k}, \bar{x_{k}})\\
    \end{aligned}
\right.
$$

\noindent 
is a system of formulae.   Consider a structure ${\cal M}$ with the 
domain $M$, interpreting all symbols in 
$\varphi_{i}$ except $R_{j}$ and $\bar{x}$.  Then $\varphi_{i}$ defines a mapping   
$\Phi_{\varphi_{i}}: 2^{M^{r_{1}}} \times \ldots 2^{M^{r_{k}}} \rightarrow 2^{M^{r_{i}}}$, where all $r_{j}$
are arities of $R_{j}$,  as follows:  
$
\Phi(P_{1}, \ldots, P_{k}) = \{\bar{a} \mid ({\cal M} \models \varphi(P_{1}, \ldots, P_{k},  \bar{a}) \}
$.  Definitions of all mentioned fixed points naturally generalize to the case of simultaneous iteration 

\[R^{0}_{i} = \emptyset\]

\[R^{j+1}_{i} = \Phi_{\varphi_{i}}(R^{j}_{1}, \ldots R^{j}_{k}).\] 

The formula $[ IND \;\; R_{i} : S]\bar{t}$ is true for a tuple of terms $\bar{t}$ if its interpretation belongs to $i$-th component 
$R_{i}^{\infty}$ of the corresponding simultaneous fixed  point. For all mentioned logics, simultaneous induction can be eliminated and equivalent formulae with simple induction can be produced \cite{FM99,Kreutzer}.

\subsection{First-order temporal logic} 
%

The language $\mT\mL$ of first order temporal logic over the natural 
numbers is constructed in the standard way
from a classical (non-temporal) first order language $\mL$ and 
a set of future-time temporal 
operators `$\sometime$' 
(\emph{sometime}), `$\always$' (\emph{always}), `$\next$' (\emph{in the next 
moment}), 
`$\until$'(\emph{until}).  


Formulae in $\mT\mL$ are interpreted in \emph{first-order temporal
structures} of the form $\mathfrak M = \auf D,\mI \zu,$ where $D$ is a
non-empty set, the \emph{domain} of $\mathfrak M$, and $\mI$ is a
function associating with every moment of time $n \in \mathbb N$ an
interpretation of predicate, function  and constant symbols of $\mL$ over $D$.
First-order (nontemporal) structures corresponding to each point of
time will be denoted $\gM_n = \auf D,\mI(n) \zu$. 

Intuitively, the interpretations of $\mT\mL$-formulae are sequences of
first-order structures, or \emph{states} of $\gM$, such as $\gM_0,
\gM_1, \ldots, \gM_n \dots.$ 

An \emph{assignment} in $D$ is a function
$\ga$ from the set $\mL_{\bf{v}}$ of individual variables of $\mL$ to
$D$.  If $P$ is a predicate symbol then $P^{\mI(n)}$ (or simply
$P^{n}$ if $\mI$ is understood) is the interpretation of $P$ in the
state $\gM_n$.  

We require that (individual) variables and constants
of $\mT\mL$ are \emph{rigid}, that is neither assignments nor
interpretations of constants depend on the state in which they are
evaluated.

The \emph{satisfaction} relation $\gM_n \gma \varphi$ (or simply $n \gma \varphi$,
if $\gM$ is understood ) in the structure $\gM$ for the assignment
$\ga$ is defined inductively in the usual way under the following
semantics of 
temporal operators:
\begin{itemize}
\item[] $$
        \begin{array}{lcl}
         n \gma \next \varphi &\mbox{iff}& n+1 \gma \varphi\\
         n \gma \sometime \varphi &\mbox{iff}& \hbox{\rm there is}\  m \geq n\ \hbox{\rm
         such that}\ m \gma \varphi\\
         n \gma \always \varphi &\mbox{iff}& m \gma \varphi\ \hbox{\rm for
         all}\ m \geq n\\
         n \gma \varphi \until \psi &\mbox{iff}& \ \mbox{there is}\ m \geq n 
         \mbox{ ~such that~ } m \gma \psi \mbox{ ~and~ } \\
         & & k \gma \varphi           \mbox{ ~for every~ }  n \leq k < m
     \end{array} $$
\end{itemize}
%

%

Let $\gM$ be a temporal structure and $\psi(\bar{x})$ be a temporal formula 
  with $\bar{x}$ only free variables 
and $|\bar{x}| = k$. Then  $\psi(\bar{x})$ defines a $k$-ary relation $P$ on 
$\gM_0$ as follows: $P(\bar{a}) \leftrightarrow \gM_0 \gma \psi(\bar{x})$ where 
$\ga : \bar{x} \mapsto \bar{a}$.

%



\section{Logic with temporally accessible iteration}~\label{FO+TAI}

In all variants of inductive logics we have discussed in the previous
section, the semantics of fixed-point construction can be defined in terms of
iteration of operators, associated with some formulae. 
In this section we described a logic which generalize and subsume all these
logics. The idea is simple:  instead of defining a particular fixed-point 
construct we allow arbitrary iterations of operators defined by formulae. 
These iterations when evaluated over a structure give rise to the sequences of
relations  over that structure. Then we allow first-order temporal logic
machinery to access these sequences of relations (temporal structures)  and 
define new relations in terms of these sequences.

The syntax of $FO+TAI$ (\emph{first-order logic with temporally accessible
iterations})  
extends the standard syntax of first-order logic with the following 
construct.  Let $\varphi(R,\bar{x})$ be  a formula with free individual variables 
$\bar{x} = x_1, \ldots, x_k$ and free predicate variable $R$ of arity $k$. 
Let $\psi(\bar{z})$ be a first-order \emph{temporal formula} ($\mT\mL$-formula) 
with  free individual variables $\bar{z} = z_{1}, \ldots, z_{m}$.  

Then  

\[
\tau := [\psi(\bar{z})][I_{R,\bar{x}}\varphi]\bar{t}
\]

\noindent
is also formula, where $\bar{t}$ is a tuple of terms of the same length as
$\bar{z}$. The free variables of $\tau$ are the free variables occurring in 
$\bar{t}$ and  the free variables of $\psi$ and $\varphi$ other than 
$\bar{z}$ and $\bar{x}$, respectively. The semantics of this construct is defined as follows.

Let ${\cal M}$ be the structure with the domain $M$ and interpretations of all
predicate and functional symbols in $M$, which will denote by $P^{M}$ and
$f^{M}$.  Let $\ga$ be assignment providing an
interpretation of free variables of $\varphi$ and $\psi$ im $M$. 
Consider the iteration $R^{0} = \emptyset$ and 
$R^{i+1} = \Phi_{\varphi}(R^{i})$. It gives rise to the temporal structure 
$\gM = \gM_0, \ldots, \gM_i, \ldots$, where every $\gM_i$  is a structure ${\cal
M}$ extended by an interpretation of $R$ by $R^{i}$. In particular 
$\gM_0$  is ${\cal M}$  augmented with empty interpretation of $R$. Let $P$ is an $m$-ary relation 
defined by $\psi(\bar{z})$ on
$\gM_0$ (i.e on M). Then for any tuple $\bar{a} \in M^{m}$, $M \models [\psi(\bar{z})][I_{R,\bar{x}}\varphi]\bar{a}$     
iff $\bar{a} \in P$. As in other fixed point logics, we also allow simultaneous iteration formulae, i.e. the formulae of 
the form $\tau := [\psi(\bar{z})][I : S] \bar{t}$ where

$$
S:= 
\left\{
    \begin{aligned}
     R_{1}(\bar{x_{1}}) \leftarrow \varphi_{1}(R_{1}, \ldots, R_{k}, \bar{x_{1}})\\
     \vdots \\
     R_{k}(\bar{x_{k}})  \leftarrow \varphi_{1}(R_{1}, \ldots, R_{k}, \bar{x_{k}})\\
    \end{aligned}
\right.
$$

is a system of formulae.   
%
Simultaneous iteration 

\[R^{0}_{i} = \emptyset\]

\[R^{j+1}_{i} = \Phi_{\varphi_{i}}(R^{j}_{1}, \ldots R^{j}_{k})\] 

induces a temporal structure  
$\gM = \gM_0, \ldots, \gM_i, \ldots$, where every $\gM_j$  is a structure ${\cal
M}$ extended by interpretation of $R_{i}$ by $R^{j}_{i}$.

Let $P$ is an $m$-ary relation 
defined by $\psi(\bar{z})$ on
$\gM_0$ (i.e on M). Then for any tuple $\bar{a} \in M^{m}$, 
$M \models [\psi(\bar{z})][I : S]\bar{a}$  iff $\bar{a} \in P$.


\begin{proposition}

FO+TAI with simultaneous iteration has the same expressive power as FO+TAI with singular iteration.  
\end{proposition} 

{\bf Proof} (hint). The proof proceed  by standard argument based on faithful modelling of simultaneous 
iteration  by a single iteration of higher-dimensional joint operator. Full details of such modelling 
(for LFP, IFP, PFP) can be found in \cite{FM99}. 


\section{FO+TAI vs other fixed point logics}\label{sec:others}

In this section we show that FO+TAI subsumes many  fixed point 
logics. We start with classical fixed point constructs.

\subsection{Least Fixed Point}
Translation of LFP construct in FO+TAI follows literally a description of the least fixed point as a limit - least fixed point consists of precisely those tuples which \emph{eventually} appear in approximations:  

\begin{description}
\item[LFP:]  $[LFP_{R, \bar{x}} \varphi(R,\bar{x})]\bar{t} \;\;\;\Leftrightarrow \;\;\; [\sometime R(\bar{z})][I_{R, \bar{x}}\varphi(R,\bar{x})]\bar{t}$  
\end{description}

Here we assume of course that $R$ is positive in $\varphi(R, \bar{x})$.

\subsection{Inflationary Fixed Point} 

Similarly to the case of LFP we have for Inflationary Fixed Point the following definition: 

\begin{description}
\item[IFP:]  $[IFP_{R, \bar{x}} \varphi(R, \bar{x})]\bar{t} \;\;\; \Leftrightarrow \; \; \; 
[\sometime R(\bar{z})][I_{R, \bar{x}} (R(\bar{x}) \lor \varphi(R,\bar{x}))]\bar{t}$ 
\end{description}

\subsection{Partial Fixed Point} 

The following definition 

\begin{description}
\item[PFP:] $[PFP_{R, \bar{x}} \varphi(R,\bar{x})]\bar{t} 
\;\;\;\Leftrightarrow \;\;\; [\sometime (R(\bar{z}) \land \forall \bar{v} (R(\bar{v}) \Leftrightarrow 
\next R(\bar{v}))  )][I_{R, \bar{x}}\varphi(R,\bar{x})]\bar{t}$  
\end{description}

\noindent 
says that Partial Fixed Point consists of the tuples satisfying two conditions: 1) a tuple should appear at some stage $i$  of iterations, and furthermore  2) approximations at the stages $i$ and $i+1$ should be the same.

\subsection{General PFP}\label{subsec:gen}

In \cite{Kreutzer} an alternative semantics for PFP has been defined under the name general PFP. 
Unlike the standard PFP general PFP generalizes easily to infinite structures and having the same expressive power as  standard PFP over finite structures provides sometimes with more concise and  natural equivalent formulae. As we mentioned in the Introduction, in this paper we consider only finite structures semantics and for this case definition of general PFP  is as follows. Let $\Phi$ is an operator defined by an arbitrary formula $\varphi(R,\bar{x})$. Consider the iteration: 
\begin{itemize}
\item  $R_{0} = \emptyset$
\item  $R_{i+1} = \Phi(R_{i})$ 
\end{itemize}  
 
Then \emph{general partial fixed point} of $\Phi$ is defined \cite{Kreutzer} as a set of tuples which occur in \emph{every stage of the first cycle} in the sequence of stages. . As noticed in \cite{Kreutzer},  in general,   this definition is not equivalent to saying that the fixed point consists of those tuples which occur at \emph{all stages starting from some stage}. Non-equivalence 
of two definitions can be established if transfinite iteration is allowed.  Since we consider the iteration over finite structures only, a cycle,  that is a sequence $R_{i}, \ldots, R_{j}$ with $R_{i} = R_{j}$,  will necessarily appear at some \emph{finite stages} $i$  and $j$. Based in that,  for the case of finite structures we have the following equivalent definition of PFPgen in terms of FO+TAI: 

\begin{description}
\item[PFPgen:] $[PFPgen_{R, \bar{x}} \varphi(R,\bar{x})]\bar{t} 
\;\;\;\Leftrightarrow \;\;\; [\sometime \always R(\bar{z})][I_{R, \bar{x}}\varphi(R,\bar{x})]\bar{t}$
\end{description}

The definition says that general PFP consists of those tuples which occur at all \emph{finite} stages starting from some stage of iteration.

\subsection{Anti-monotone induction} 

Let  $\Phi_{\varphi}$ be an operator associated with a formula $\varphi(P,\bar{x})$. It may turn out that this operator
is \emph{anti-monotone}, that is $P \subseteq P' =>  \Phi_{\varphi}(P') \subseteq \Phi_{\varphi}(P)$. Syntactical
condition which  entails anti-monotonicity  is that the predicate variable $P$ has only negative occurrences in
$\varphi(P, \bar{x})$.  As before consider the iteration $R_{0} = \emptyset$, $R_{i+1} = \Phi(R_{i})$.

An interesting analogue of classical Knaster-Tarski result holds \cite{Yablo,Fitting}: the above iteration of
anti-monotone operator converges to a pair of oscillating points $P$ and $Q$ that is $Q = \Phi(P)$ and $P = \Phi(Q)$.
What is more, one of the oscillating points is a least fixed point $\mu$ and another is the greatest fixed 
point $\nu$ of the
monotone operator $\Phi^{2}$ (where $\Phi^{2}(X) = \Phi(\Phi(X)))$.) 

One may extend then the first-order logic  with suitable oscillating points constructs 
$[OP^{\mu}_{R, \bar{x}} \varphi(R,\bar{x})]\bar{t}$ and $[OP^{\nu}_{R, \bar{x}} \varphi(R,\bar{x})]\bar{t}$ for
$\varphi(R,\bar{x})$ negative in $R$, with obvious semantics. Because of the definability
of oscillating points as the fixed points of $\Phi^{2}$, first order logic extended with these
constructs is no more expressive than  FO+LFP and therefore than FO+TAI. What is interesting 
here is that FO+TAI
allows to define oscillating points directly, not referring to LFP construct. 
Here it goes.  
For the greater of two oscillating points we have 
$[OP^{\nu}_{R, \bar{x}} \varphi(R,\bar{x})]\bar{t} \Leftrightarrow [\psi^{\nu}(R)][I_{R, \bar{x}}\varphi(R,\bar{x})]\bar{t}
$

\noindent 
where  $\psi^{\nu}(R)$ is the temporal formula $\sometime (R(\bar{z}) 
\land \forall \bar{y} (R(\bar{y}) \leftrightarrow \next\next R(\bar{y})) \land (\exists \bar{y} (R(\bar{y}) \land
\next \neg R(\bar{y})) \lor \forall \bar{y} (R(\bar{y}) \leftrightarrow \next R(\bar{y}) )  )  )$

Similarly, $[OP^{\mu}_{R, \bar{x}} \varphi(R,\bar{x})]\bar{t} \Leftrightarrow [\psi^{\mu}(R)][I_{R, \bar{x}}\varphi(R,\bar{x})]\bar{t}
$

\noindent 
where  $\psi^{\mu}(R)$ is the temporal formula $\sometime (R(\bar{z}) 
\land \forall \bar{y} (R(\bar{y}) \leftrightarrow \next\next R(\bar{y})) \land (\exists \bar{y} (\neg R(\bar{y}) \land
\next R(\bar{y})) \lor \forall \bar{y} (R(\bar{y}) \leftrightarrow \next R(\bar{y}) )  )  )$

\subsection{Some variations}

In the above FO+TAI definition  for LFP it is assumed that 
$\varphi(R, \bar{x})$ is positive  in 
$R$. If we consider the same right-hand side definition 
$[\sometime R(\bar{z})][I_{R, \bar{x}}\varphi(R,\bar{x})]\bar{t}$ for not necessarily positive (and monotone) 
$\varphi(R,\bar{x})$  than we get definition of an  operator which does not have direct analogue in 
standard fixed-point logics and may be considered as a variation of PFP, which we denote by $PFP^{\cup}$.
Similarly, one can define: $[PFP^{\cap}_{R, \bar{x}} \varphi(R,\bar{x})]\bar{t} 
\;\;\;\Leftrightarrow \;\;\; [\always R(\bar{z})][I_{R, \bar{x}}\varphi(R,\bar{x})]\bar{t}$ 

\noindent 
It has  turned out though that both    
$PFP^{\cup}$ and $PFP^{\cap}$ are easily definable by simultaneous partial fixed-points, for details see Theorem~\ref{thm:TAI=PFP}.

If in definition of PFPgen we swap temporal operators we get a definition of what can be called Recurrent Fixed Point (RFP)\footnote{Notice, than in general, and similarly to $PFPgen$,  neither of $PFP^{\cup}$, $PFP^{\cap}$, RFP define  
fixed points of any natural operators. 
But we follow \cite{Kreutzer} and preserve  the name ``fixed points'' and FP
in abbreviations.}: 

\begin{description}
\item[RFP:] $[RFP_{R, \bar{x}} \varphi(R,\bar{x})]\bar{t} 
\;\;\;\Leftrightarrow \;\;\; [\always \sometime R(\bar{z})][I_{R, \bar{x}}\varphi(R,\bar{x})]\bar{t}$
\end{description}

Again it is not difficult to demonstrate that $RFP$ is definable in terms of either PFP or PFPgen. 


\section{ID-logic of non-monotone induction}~\label{sec:ID}

In \cite{DT} a logic of non-monotone definitions has been introduced.  Motivated by well founded semantics for logic programming, ID-logics formalises   non-monotone, in general, inductive definitions of the form $P \leftarrow \varphi(P)$ where predicate variable $P$ may have both positive and negative occurrences in $\varphi(P)$. It subsumes and generalizes both monotone and anti-monotone inductions.
The main point in the definition of ID-logic is a semantics given to non-monotone inductive definition  which we present here in a operator form\footnote{In \cite{DT} inductive definitions of ID-logic are presented not by operators, but by special formulae called \emph{definitions}. The difference is purely syntactical and insignificant for our discussion   here.}. Similarly to already discussed fixed point extensions, the syntax of ID-logic (this version we call FO+ID) extends the standard syntax of 
first-order logic with  the following construct. Let $\varphi(P,\bar{x})$ be a formula with free individual variables
$\bar{x} = x_{1}, \ldots, x_{k}$ and free predicate variable $P$. Then $\rho := [ID_{P, \bar{x}} (P(\bar{x}) \leftarrow \varphi(P, \bar{x}))]\bar{t}$ is
also formula. 
Now we explain semantics of this construct in terms of FO+TAI, showing thereby that FO+ID is also subsumed by FO+TAI. 
Since $\varphi(P,\bar{x})$ may have both negative and positive occurrences of $P$ the iteration of the operator $\Psi_{\varphi}$ applied to the empty interpretation of $P$ will not necessary converge to a  fixed  point. In the  semantics adopted in FO+ID, the extension of defined predicate is obtained as a \emph{common limit} of iteratively computed 
lower and upper bounds (if it exists). Introduce two new auxiliary predicate variables $P_{l}$ and $P_{u}$, with the intended meaning to be \emph{lower} and \emph{negated upper} approximations for the defined predicate. Further, denote by $\varphi(P_{l})$, respectively,  by $\varphi(\neg P_{u})$ the result of replacement of all negative occurrences of $P$ in $\varphi(P,\bar{x})$ with $P_{l}$, resp. with $\neg P_{u}$. All positive occurrences of $P$ remains unaffected in both cases. Consider then the following definition of the step of simultaneous iteration:

$$
S:= 
\left\{
    \begin{aligned}
     P_{u}(\bar{y}) \leftarrow \neg [LFP_{P, \bar{x}} (\varphi(P_{l}))]\bar{y}\\
    \vdots \\
     P_{l} (\bar{y})  \leftarrow [LFP_{P, \bar{x}} (\varphi(\neg P_{u}))]\bar{y}\\
    \end{aligned}
\right.
$$

Since both 
$\varphi(P_{l})$ and $\varphi(\neg P_{u})$ are positive in $P$ the least fixed point operators in the right hand sides of the definitions are well-defined. 

Starting with $P^{0}_{l} = P^{0}_{u}  = \emptyset$ and iterating this definition one gets a sequences of lower and \emph{negated} upper approximations $P^{i}_{l}$ and $P^{i}_{u}$. If the lower and upper approximations converge to the same limit, i.e. $P^{\infty}_{l} = \neg P^{\infty}_{u}$ then by definition \cite{DT} this limit is taken as the predicate defined by the above ID-construct. Summing up,  the FO+ID formula $\rho$ shown above is equivalent  to the  following formula of FO+TAI:

\[
[\sometime (P_{L}(\bar{x}) \land \forall \bar{y} (P_{l}(\bar{y}) \leftrightarrow \neg P_{u}(\bar{y}))][I:S^{\ast}]\bar{t}
\] 

where $S^{\ast}$ is obtained of the above $S$ by translation of the right hand side parts of S  into FO+TAI.


\section{Expressive power}\label{sec:expr} 

We have seen in previous sections that FO+TAI is very expressive logic and subsumes many other fixed-point logics,
including most expressive (among mentioned) FO+PFP (and FO+PFPgen). The natural question is 
whether FO+TAI is more expressive than FO+PFP? In this section we answer this question negatively and show that for any
formula in FO+TAI one can effectively produce an equivalent (over finite structures) FO+PFP formula.

\begin{theorem}\label{thm:TAI=PFP}
For every formula $\tau := [\psi][I_{R,\bar{x}}\varphi]\bar{t}$ of FO+TAI there is an equivalent 
formula $\tau^{\ast}$ of FO + PFP
\end{theorem}

{\bf Proof} The formula $\tau^{\ast}$ equivalent to a $\tau$ is build by induction on the construction of $\tau$. 
Correctness of the proposed translation $\tau \mapsto \tau^{\ast}$ is established by induction along the construction. 
Correctness of the base case and induction steps follows by routine check of definitions.   

If $\tau := [\psi(\bar{z})][I_{R, \bar{x}}\varphi]\bar(t)$ with $[\psi]$ non-temporal formula 
then 

\[ [\tau]^{\ast} := \psi(\bar{z})_{\mid R \leftarrow \emptyset, \bar{z} \leftarrow \bar{t}}
\]

where $R \leftarrow \emptyset$ means substitute all occurrences of $R$ in $\psi$ with $\exists x\not=x$ and 
$ \bar{z} \leftarrow \bar{t}$ substitute $\bar{t}$ into $\bar{z}$. 

 The cases of boolean connectives and quantifiers in the head and body of the formula are straightforward. 

\begin{itemize}

\item $([\psi_{1} \land \psi_{2}] [I_{R, \bar{x}} \varphi] \bar{t})^{\ast} = 
([\psi_{1}] [I_{R, \bar{x}} \varphi] \bar{t})^{\ast} \land ([\psi_{2}] [I_{R, \bar{x}} \varphi] \bar{t})^{\ast}$

\item $([\psi_{1} \lor \psi_{2}] [I_{R, \bar{x}} \varphi] \bar{t})^{\ast} = 
([\psi_{1}] [I_{R, \bar{x}} \varphi] \bar{t})^{\ast} \lor ([\psi_{2}] [I_{R, \bar{x}} \varphi] \bar{t})^{\ast}$

\item $([\neg \psi] [I_{R, \bar{x}} \varphi] \bar{t})^{\ast} = \neg ([\psi] [I_{R, \bar{x}} \varphi] \bar{t})^{\ast}$

\item  $([\exists y. \psi] [I_{R, \bar{x}} \varphi] \bar{t})^{\ast} = \exists y. ([\psi] [I_{R, \bar{x}} \varphi]
\bar{t})^{\ast}$

\item  $([\forall y. \psi] [I_{R, \bar{x}} \varphi] \bar{t})^{\ast} = \forall y. ([\psi] [I_{R, \bar{x}} \varphi] 
\bar{t})^{\ast}$

\item $(\varphi \land \psi)^{\ast}  = \varphi^{\ast} \land \psi{\ast}$

\item $(\neg \varphi)^{\ast} = \neg \varphi^{\ast}$

\item $(\forall y \; \varphi)^{\ast} = \forall y \varphi^{\ast}$

\end{itemize}

\begin{itemize}

\item If $\tau =  [\sometime \psi(\bar{z})][I_{R, \bar{x}}\varphi(R,\bar{x})]\bar{t}$ then $\tau^{\ast} := [PFP \;\;Q, \bar{v}: S] \bar{t}$ where

$$
S:= 
\left\{
    \begin{aligned}
     R(\bar{x}) \leftarrow (\varphi_(R, \bar{x}))^{\ast}\\
     \vdots \\
     Q (\bar{v})  \leftarrow Q (\bar{v}) \lor  ([\psi(\bar{z})][I_{R, \bar{x}}\varphi(R,\bar{x})]\bar{v})^{\ast}\\
    \end{aligned}
\right.
$$

\item The case of $\always$-modality as the main connective in the head of iteration is reduced to the case of $\sometime$
modality: $([\always  \psi][I_{R, \bar{x}}\varphi(R,\bar{x})]\bar{t})^{\ast} = 
([\neg \sometime \neg    \psi][I_{R, \bar{x}}\varphi(R,\bar{x})]\bar{t})^{\ast}$

\item If $\tau =  [\next \psi(\bar{z})][I_{R, \bar{x}}\varphi(R,\bar{x})]\bar{t}$ then $\tau^{\ast} := [PFP \;\;Q, \bar{x}: S] \bar{t}$ where

$$
S:= 
\left\{
    \begin{aligned}
     R(\bar{x}) \leftarrow (\varphi(R, \bar{x}))^{\ast}\\
    \vdots \\
     Q (\bar{x})  \leftarrow [(\varphi(R, \bar{x})^{\ast}]^{2}\\
    \end{aligned}
\right.
$$

\item If $\tau = [(\psi_{1} \until \psi_{2})(\bar{z})][I_{R,\bar{x}} \varphi(R,\bar{x})] \bar{t}$
then $\tau^{\ast} := [PFP \;\; Q \bar{x} : S] \bar{t}$ where 

$$
S :=
\left\{
   \begin{aligned}
    R(\bar{x}) \leftarrow (\varphi(R, \bar{x}))^{\ast}\\
    P (\bar{z}) \leftarrow P(\bar{z}) \lor \neg \psi_{1}(\bar{z})\\
    Q (\bar{z}) \leftarrow Q(\bar{z}) \lor (\neg P(\bar{z} \land \psi_{2}(\bar{z})\\
   \end{aligned}
\right.
$$

\end{itemize}

\subsection{Temporally accessible monotone induction}

What happens if we apply temporal logic based access to the iteration steps of monotone induction? Will the 
resulting logic be more expressive than the logic of the monotone induction? Negative answer is given by the following theorem.  

\begin{theorem}
For every formula $\tau := [\psi][I_{R,\bar{x}}\varphi]\bar{t}$ of FO+TAI with $\varphi$ positive in $R$ 
there is an equivalent formula $(\tau)^{\ast}$ of FO + LFP
\end{theorem} 

\noindent {\bf Proof} 
%
The translation here uses the \emph{stage comparison theorem} of 
Moschovakis \cite{Mos74}.  With any monotone map $\Phi_{\varphi}$ of arity $k$ 
defined by a positive in $R$ formula  and a structure with finite domain $M$ on can
associate a rank function $\mid \;\; \mid_{\Phi}: M^{k} \rightarrow N \cup \{\infty\}$ 
which 
%
when applied to  any tuple of elements $\bar{a} \in M^{k}$ yeilds the least
number $n$ such that $\bar{a} \in \Phi^{n}(\emptyset)$ if such $n$ exists and
$\infty$ otherwise, i.e. when $\bar{a} \not\in \Phi^{\infty}$  

Stage comparison relation $\le_{\Phi}$ defined as $\bar{a} \le_{\Phi} \bar{b} \Leftrightarrow$  $\bar{a}, \bar{b} \in
\Phi_{\phi}(\emptyset)$ and $\mid \bar{a} \mid \le \mid \bar{b} \mid$.

\begin{theorem}
For any $LFP_{\varphi}$ operator associated with a first-order formula $\varphi(P,\bar{x})$ positive in $P$ the stage comparison relation $\le_{\varphi}$ is definable in FO+LFP uniformly over all finite structures. 
\end{theorem}

The stage comparison relation can be used then to simulate time in modelling temporal access to the iteration steps within FO+LFP. As above, the translation is defined by induction on formula structure. We present here only  translation of 
$[\psi(\bar{z})][I_{R,\bar{x}}\varphi]\bar{t}$ where $\psi(\bar{z})$ is a temporal formula and $\varphi$ is in FO+LFP. 


For a formula $[\psi(\bar{z})][I_{R,\bar{x}}\varphi]\bar{t}$, define translation of its temporal header 
$\psi(\bar{z})$ in the \emph{context of iteration} $[I_{R,\bar{x}}\varphi]$, to a formula in FO+LFP.   
Translation is indexed by either a constant $s$ ( from $start$)  or a tuple of variables of the same length as the arity of the predicate used in iteration definition, i.e. of $R$:



\begin{itemize}
\item $[P(\bar{x})]^{s} := P(\bar{x})$ (for any predicate $P$).
\item $[R(\bar{x})]^{\bar{u}} := \bar{x} \le_{\varphi} \bar{u} \land R(\bar{x})$ (for the iteration predicate $R$)
\item $[P(\bar{x})]^{\bar{u}} := P(\bar{x})$ (for any predicate $P$ other than iteration predicate)   
\item $[\rho \land \tau]^{s} := [\rho]^{s} \land [\tau]^{s}$
\item $[\rho \land \tau]^{\bar{u}} := [\rho]^{\bar{u}} \land [\tau]^{\bar{u}}$
\item $[\neg \rho]^{s} := \neg [\rho]^{s}$
\item $[\neg \rho]^{\bar{u}} := \neg [\rho]^{\bar{u}}$ 
\item $[\forall x \rho]^{s} := \forall x [\rho]^{s}$
\item $[\forall x \rho]^{\bar{u}} := \forall x [\rho]^{\bar{u}}$ 
\item $[\next \tau]^{s} := \exists \bar{u} (\varphi(\bar{u})  \land [\tau]^{\bar{u}})$
\item $[\next \tau]^{\bar{u}} := \exists \bar{u'}(next_{\varphi}(\bar{u},\bar{u}') \land [\tau]^{\bar{u}'})$
\item $[\sometime \tau]^{s} := \exists \bar{u} [LFP_{R, \bar{x}} \varphi]\bar{u} \land  [\tau]^{\bar{u}}$
\item $[\sometime \tau]^{\bar{u}} := \exists (\bar{u}' (\bar{u} \le_{\varphi} \bar{u}') \land [\tau]^{\bar{u}'})$
\item $[\rho \until \tau]^{s} := \exists \bar{u} ([LFP_{R, \bar{x}} \varphi]\bar{u} \land 
[\tau]^{\bar{u}}) \land \forall \bar{u}' (\bar{u}' <_{\varphi} \bar{u}) \rightarrow 
[\rho]^{\bar{u}'}$
\item $[\rho \until \tau]^{\bar{u}} := \exists \bar{u}' (\bar{u} \le_{\varphi} \bar{u}' \land 
[\tau]^{\bar{u}'}) \land \forall \bar{u}'' (\bar{u} \le_{\varphi} \bar{u}'' <_{\varphi} \bar{u}') \rightarrow 
[\rho]^{\bar{u}''}$
\end{itemize}

Now to get a formula in FO+LFP equivalent to $[\psi(\bar{z})][I_{R,\bar{x}}\varphi]\bar{t}$ we take 
translation $[\psi(\bar{z})]^{s}$ in the context of $[I_{R,\bar{x}}\varphi]\bar{t}$.


\section{Concluding remarks}\label{sec:conclusion}  
We proposed in this paper the logic with temporally accessible iteration which provides the simple unifying framework for studying logics with inductive fixed point operators. Obvious next step is to extend the semantics to the case of infinite structures. Also of interest are modifications of FO+TAI with branching time access to incorporate non-deterministic inductive definitions
\cite{DawarGurevich} and modal variants of the logic. Probably most interesting applications FO+TAI may find in formal analysis of revision theory in the spirit of recent conceptual idea  
\cite{Lowe:revision} to analyse ``the nonmonotonic process by looking at the behaviour of interpretations under revision rules''.


\end{document}